\documentclass{llncs}

\usepackage[dvips]{epsfig}

\title{Variable and Value Ordering When Solving Balanced Academic Curriculum Problems\thanks{This work has been partially funded by the National Science and Technology  Fund of Chile, FONDECYT, under grant No. 1010121.}}

\author{Carlos Castro \& Sebasti\'an Manzano}

\institute{{\small Departamento de Inform\'atica, Universidad T\'ecnica Federico Santa Mar{\'\i}a} \\
{\small Avenida Espa\~na 1680, Casilla 110-V, Valpara{\'\i}so, Chile}\\
{\small {\tt \{ccastro,smanzano\}@inf.utfsm.cl}}}
\date{}

\begin{document}
\maketitle

\begin{abstract}
In this paper we present the use of Constraint Programming for solving
balanced academic curriculum problems. We discuss the important role
that heuristics play when solving a problem using a constraint-based
approach. We also show how constraint solving techniques allow to very
efficiently solve combinatorial optimization problems that are too
hard for integer programming techniques.
\end{abstract}

\begin{quote}
{\small {\bf Key words}: Combinatorial Optimization, Variable Ordering, Value Ordering.}
\end{quote}

\section{Introduction}

A key factor to take into account when evaluating the academic success
of students is the academic load they have in each academic
period. The usual way to measure this load is to assign, to each
course, a number of credits representing the amount of effort required
to successfully follow the course. In this way, the academic load of
each period is given by the sum of the credits of each course taken in
the period. Generally, some explicit restrictions are imposed when
developing a curriculum. For example, a maximum load per period could
be allowed in order to prevent overload, and some precedence
relationships could be established among some courses. Assuming that a
balanced load favors academic habits and facilitates the success of
students, since $1994$ we have been involved in designing balanced
academic curricula at the Federico Santa Mar{\'\i}a Technical
University.

The problem of designing balanced academic curricula consist in
assigning courses to periods in such a way that the academic load of
each period will be balanced, i.e., as similar as possible. In this
work, we consider as academic load the notion of credit that
represents the effort in time needed to successfully follow the
course. In this work, we concentrate on the Informatics careers
offered by the Federico Santa Mar{\'\i}a Technical University at
Valpara{\'\i}so. A first attempt in this direction was done on the
curriculum of a four-year career ($8$ academic periods) considering
$48$ courses~\cite{Vergara:IEI:1994}. In that work, Integer
Programming techniques allowed to solve only $6$ academic
periods. However, it was not possible to solve the complete
model. Considering the success of Constraint Programming for solving
combinatorial search problems, we were interested in using this
technique as an alternative.

This paper is organized as follows: section \ref{problem} describes
the combinatorial optimization problem we are interested in. Section
\ref{pl_model} presents an Integer Programming model for the balanced
academic curriculum problem. In section \ref{cp_model}, we model the
same problem using a constraint-based approach. In section
\ref{results}, we present experimental results when solving the
problem using both approaches. Finally, in section \ref{conclusions},
we conclude the paper and give some perspectives for further works.

\section{The Balanced Academic Curriculum Problem}
\label{problem}

In this work, we concentrate on three particular instances of the
balanced academic curriculum problem: the three Informatics careers
offered by the Federico Santa Mar{\'\i}a Technical University. As a
general framework we consider administrative as well as academic
regulations of this university.

\begin{description}

\item[Academic Curriculum] An academic curriculum is defined by a set
of courses and a set of precedence relationships among them.\\

\item[Number of periods] Courses must be assigned within a maximum
number of academic periods.\\

\item[Academic load] Each course has associated a number of credits or
units that represent the academic effort required to successfully
follow it.\\

\item[Prerequisites] Some courses can have other courses as
prerequisites.\\

\item[Minimum academic load] A minimum amount of academic credits per
period is required to consider a student as full time.\\

\item[Maximum academic load] A maximum amount of academic credits
per period is allowed in order to avoid overload.\\

\item[Minimum number of courses] A minimum number of courses per
period is required to consider a student as full time.\\

\item[Maximum number of courses] A maximum number of courses per
period is allowed in order to avoid overload.

\end{description}

\newpage

\section{Integer Programming Model}
\label{pl_model}

In this section, we present an Integer Programming model for the
balanced academic curriculum problem.

\begin{itemize}

\item Parameters

Let

\begin{description}

\item[$m$]: Number of courses

\item[$n$]: Number of academic periods

\item[$\alpha_i$]: Number of credits of course $i$; $\forall i = 1
,\dots , m$

\item[$\beta$]: Minimum academic load allowed per period

\item[$\gamma$]: Maximum academic load allowed per period

\item[$\delta$]: Minimum amount of courses per period

\item[$\epsilon$]: Maximum amount of courses per period

\end{description}

\item Decision variables

Let



 \begin{displaymath}
   x_{ij} = \left\{ \begin{array}{ll}
                     1 & $if course $i$ is assigned to period $j ; \forall i = 1 , \dots , m \; \forall j = 1 ,\dots , n \\
                     0 & $otherwise$
                    \end{array}
            \right.
  \end{displaymath}


 $c_j$: academic load of period $j$; $\forall j = 1, \dots , n$


 $c$: maximum academic load for all periods


\item Objective function

  \begin{displaymath}
    Min ~~ c = Max \{ c_1 , \ldots, c_n \}
  \end{displaymath}

\item Constraints

\begin{itemize}

\item The academic load of period $j$ is defined by:

  \begin{displaymath}
    c_j = \sum \limits_{i=1}^{m} \alpha_i \times x_{ij} \; \forall j = 1,\dots, n
  \end{displaymath}

\item All courses $i$ must be assigned to some period $j$:

  \begin{displaymath}
    \sum\limits_{j=1}^{n} x_{ij} = 1 \; \forall i = 1,\dots,m
  \end{displaymath}

\item Course $b$ has course $a$ as prerequisite:

  \begin{displaymath}
    x_{bj} \le \sum\limits_{r=1}^{j-1} x_{ar} = 1 \; \forall j = 2 , \dots , n
  \end{displaymath}

\item The maximum academic load is defined by:

  \begin{displaymath}
    c = Max \{c_1 , \ldots, c_n \}
  \end{displaymath}

This can be represented by the following set of linear constraints:

  \begin{displaymath}
    c_j \le c \; \forall j = 1,\dots, n
  \end{displaymath}

\item The academic load of period $j$ must be greater than or equal to
the minimim required:

  \begin{displaymath}
    c_j \ge \beta \; \forall j = 1,\dots, n
  \end{displaymath}

\item The academic load of period $j$ must be less than or equal to
the maximum allowed:

  \begin{displaymath}
    c_j \le \gamma \; \forall j = 1,\dots, n
  \end{displaymath}

\item The number of courses of period $j$ must be greater than or
equal to the minimum allowed:

  \begin{displaymath}
    \sum \limits_{i=1}^{m} x_{ij} \ge \delta \; \forall j = 1,\dots,n
  \end{displaymath}

\item The number of courses of period $j$ must be less than or equal
to the maximum allowed:

  \begin{displaymath}
    \sum \limits_{i=1}^{m} x_{ij} \le \epsilon \; \forall j = 1,\dots,n
  \end{displaymath}

\end{itemize}
\end{itemize}

\section{Constraint-Based Model}
\label{cp_model}

In this section, we present our constraint-based model using the Oz
language.

\subsection{Variables}
\label{subsec:variables_oz}

The domain constraints are the following:

\begin{center}
\begin{tabular}{lll}\hline\hline 
\textbf{Variable} &
\textbf{Description} & \textbf{Oz Implementation} \\ \hline
C       & Maximum academic load &
\texttt{\{FD.int 0\#FD.sup C\}} \\
 &  (upper bound for all periods) & \\
Assignment  & Binary vector of assignments & \texttt{\{FD.tuple assignment NK} \\
 &  course-period (size N $\times$ K) & \texttt{0\#1 Assignment\}} \\ \hline
\end{tabular}
\end{center}

\subsection{Basic constraints}

Constraints involving only one variable and that allow to narrow the
domains are:

\begin{center}
\begin{tabular}{ll}\hline\hline 
\textbf{Description}  & \textbf{Oz Implementation} \\
\hline
Maximum academic load & \texttt{C =<: Gamma}       \\
Minimum academic load & \texttt{C >=: Beta}        \\
\hline
\end{tabular}
\end{center}

\subsection{Propagators}

The academic load of each period must be less than or equal to
\texttt{C}.  The product of the \emph{load vector} by each column of
the \emph{assignment matrix} must be less than or equal to \texttt{C}.

\begin{verbatim}
      % Academic load for period J
      {For 1 K 1
       proc {$ J}
          {FD.sumC
           {Map L1N fun {$ I} {Nth   I LoadVector} end}
           {Map L1N fun {$ I} {Field I J} end}
           '=<:' C}
       end}
\end{verbatim}

Any course must be included only once. The sum of any column of the
\emph{assignment matrix} must be $1$.

\begin{verbatim}
      % Any course must be done only once
      {For 1 N 1
       proc {$ I}
          {FD.sum
           {Map L1K fun {$ J} {Field I J} end}
           '=:' 1}
       end}
\end{verbatim}

The number of courses assigned to any period must be less than or
equal to the maximum allowed.  The sum of any row of the
\emph{assignment matrix} must be less than or equal to the maximum
allowed (\texttt{Epsilon}).

\begin{verbatim}
      % Maximum number of courses in a period
      {For 1 K 1
       proc {$ J}
          {FD.sum
           {Map L1N fun {$ I} {Field I J} end}
           '=<:' Epsilon}
       end}
\end{verbatim}

Each course must be assigned after its prerequisites. If the course to
be assigned has any precedence constraint, then the simple constraint
\texttt{\{Field A 1\} =: 0} is inserted because it could not be
assigned to the first period. Then, for each ancestor, the propagators
are implemented.

\newpage

\begin{verbatim}
% Course Precedence Relationship
{ForAll Courses
 proc {$ Course}
    A P
 in
    A = {Order Course Courses}
    P = {Nth A PrecedenceRelations}
    if P \= nil then
       {Field A 1} =: 0
       {ForAll P
        proc {$ Preced}
          B
        in
          B = {Order Preced Courses}
          {For 2 K 1
           proc {$ J}
            {FD.sum
            {Map {List.number 1 J-1 1} fun {$ R} {Field B R} end}
            '>=:' {Field A J}}
           end}
        end}
     end
 end}
\end{verbatim}

\section{Experimental Results}
\label{results}

In this section, we detail the results obtained when solving the
balanced academic curriculum problem using integer programming and
constraint programming techniques. We consider three careers involving
$8$, $10$ and $12$ academic periods, respectively.

\subsection{\texttt{lp\_solve}: The Integer Programming Approach}

Several methods are available for solving Integer Programming
problems. Explicit enumeration and cutting planes algorithms provide
satisfactory results in some cases. However, Branch and Bound
algorithms are the most successful techniques for solving Integer
Programming models and are included in almost all mathematical
programming packages.

Specifically, when solving Integer Programming models that involve
only binary variables, Branch and Bound algorithms work in the
following way: the original problem $P$ is relaxed by eliminating
constraints that impose integer values for the variables. Then, the
relaxed problem is solved by using linear programming techniques and,
in case the solution does not satisfy the requirement of integer
values for the variables, two subproblems are created. The first one
corresponds to the original problem $P$ plus the additional constraint
$x=0$, and the second one corresponds to the original problem $P$ plus
the additional constraint $x=1$. The variable $x$ is called the
branching variable. Each time subproblems are created only worse
solution can be obtained and this fact is used for bounding the search
tree~\cite{NemhauserWolsey:1999}.

Table \ref{tab:lp_solve_8sem} presents the results obtained using
\texttt{lp\_solve}\footnote{{\tt lp\_solve}, a software for solving
integer linear programming models, is available free of charge at {\tt
ftp://ftp.ics.ele.tue.nl/pub/lp\_solve/}} when solving the $8$-period
problem. In this case, the optimum plan is obtained in $1460$ seconds.

\begin{table*}[ht!]
\center
\begin{tabular}{cc||cc}\hline\hline
 Solution quality & Time &  Solution quality & Time \\
  \texttt{C} {[credits]}  & [seconds] &  \texttt{C} {[credits]}  & [seconds] \\ \hline
54 & 1.69 & 33 & 9.35  \\
52 & 1.80 & 32 & 11.63 \\
50 & 2.09 & 30 & 11.94 \\
48 & 2.41 & 29 & 53.64 \\
47 & 2.97 & 27 & 54.04 \\
45 & 3.28 & 26 & 136.45 \\
44 & 3.91 & 24 & 137.08 \\
42 & 4.24 & 23 & 218.23 \\
41 & 4.92 & 21 & 218.43 \\
39 & 5.20 & 20 & 712.84 \\
38 & 6.18 & 19 & 1441.98 \\
36 & 6.53 & 18 & 1453.75 \\
35 & 9.04 & 17 (optimum) & 1459.73 \\\hline
\end{tabular}
\caption{Partial and final solutions obtained by \texttt{lp\_solve} for the $8$-period problem.}
\label{tab:lp_solve_8sem}
\end{table*}

\vspace{-6mm}

Table \ref{tab:lp_solve_10sem} presents the results obtained when
solving the $10$-period problem. In this case, we obtained no optimum
result after $5$ hours, and the last result was logged before $1700$
seconds.

\begin{table*}[ht!]
\center
\begin{tabular}{cc||cc}\hline\hline
 Solution quality & Time &  Solution quality & Time \\
  \texttt{C} {[credits]}  & [seconds] &  \texttt{C} {[credits]}  & [seconds] \\ \hline
 48 & 5.88 &  35 & 8.84 \\
 46 & 5.89 &  33 & 9.11 \\
 44 & 5.93 &  32 & 25.38 \\
 42 & 6.00 &  30 & 25.65 \\
 41 & 6.14 &  29 & 1433.18 \\
 39 & 6.33 &  27 & 1433.48 \\
 38 & 6.72 & 26 & 1626.49 \\
 36 & 7.00 &  24 (not optimum) & 1626.84 \\ \hline
\end{tabular}
\caption{Partial solutions obtained by \texttt{lp\_solve} for the $10$-period problem.}
\label{tab:lp_solve_10sem}
\end{table*}

\vspace{-6mm}

For the $12$-period problem we did not get any log after a
``turn-around-time'' of $1$ day.

\subsection{Oz: The Constraint-Based Approach}

Constraint Programming deals with optimization problems using the same
basic idea of verifying the satisfiability of a set of
constraints. Asuming one is dealing with a minimization problem, the
idea is to use an upper bound that represents the best possible
solution obtained so far. Then we solve a sequence of CSPs each one
giving a better solution with respect to the optimization
function. More precisely, we compute a solution $\alpha$ to the
problem $P$ and we add the constraint $f < \alpha(f)$ that restricts
the set of possible solutions to those that give better values for the
optimization function\footnote{Bockmayr and Kasper work on
optimization functions $f$ that always take integer values and so they
add the constraint $f \le \alpha(f) -
1$~\cite{BockmayrKasper:TR-008:1997}.}. When, after adding such a
constraint, the problem becomes unsatisfaible, the last possible
solution so far obtained represents the optimal solution.

\begin{figure*}[ht!]
  \begin{center}
    \leavevmode
    \epsfig{file=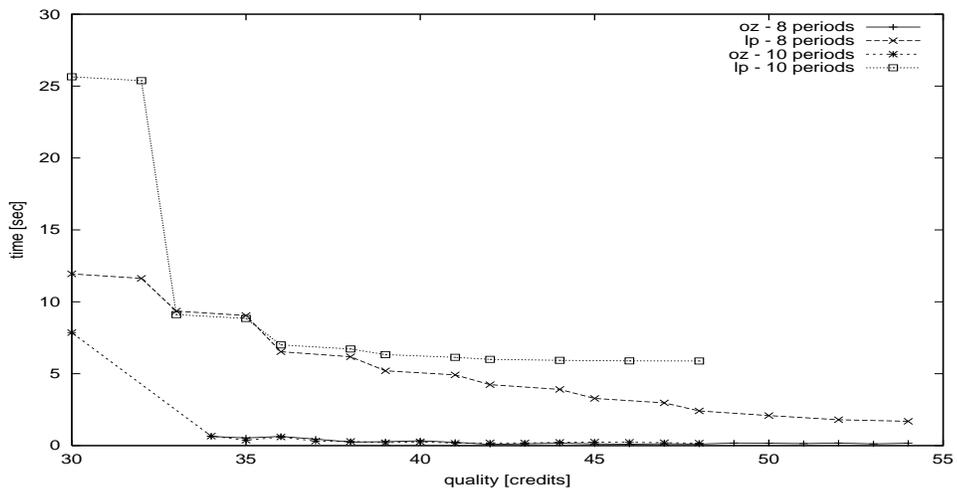,height=6.5cm,width=13cm}
    \caption{Results obtained by the first implementation of the model
    in Oz compared with the \texttt{lp\_solve} performance}
    \label{fig:oz_8-10_JI_naive}
  \end{center}
\end{figure*}

When solving our constraint-based model implemented in Oz, we first
used a naive heuristic over the \emph{assignment matrix} to find a
feasible solution. We did not have any other choice because the
representation was a binary matrix involving $n \times m$ variables
which are instantiated (with a $0$ or $1$ value) or not instantiated (with
a 0\#1 domain) at any time. Because the other heuristics implemented
by Oz (\texttt{first fail}, \texttt{split}) only apply to variables
with bigger domains (e.g. \texttt{first fail} chooses the variable
with minimal domain size) we only use the \texttt{naive} heuristic
varying the order and interpretation of the vector of assignments. The
way the program treats the assignment vector involves the
\emph{variable ordering}, and we could reverse the variable ordering
by changing a function that is used along the model:

\begin{verbatim}
      fun {Field I J} 
            Assignment.((J-1)*N + I)
      end
\end{verbatim} 

If we change the order of the matrix, then the variables are
instantiated in a left-right manner always in the assignment vector
but in a different way along the assignment matrix (interpretation of
that vector). The Oz distributor is:

\begin{verbatim}
      {FD.distribute generic(order:naive value:max) Assignment}
\end{verbatim}

The first model we tested was the same model solved by
\texttt{lp\_solve} but now implemented as a CSP problem in Oz. The
results for successive iterations reducing \texttt{C} (the upper bound
of the number of credits) are shown in figure
\ref{fig:oz_8-10_JI_naive}, having obtained solutions before $5$ hours
only for the 8- and 10-period problems.

The Oz performance was better than \texttt{lp\_solve} giving poor
quality solutions faster. But when we incremented \texttt{C} the
problem got thighter and the latter still gave solutions (see tables
\ref{tab:lp_solve_8sem} and \ref{tab:lp_solve_10sem}) while Oz either
ran out of memory\footnote{That was an Oz internal error sent to
\texttt{bugs@mozart-oz.org}. In fact, the entire search space
(considering only distribution, not propagation) is rather big
($2^{420}$ for the $10$-period problem in $42$ courses, $10$
periods).}  or did not log any solution after $5$ hours.

\begin{figure}[ht!]
  \begin{center}
    \leavevmode
    \epsfig{file=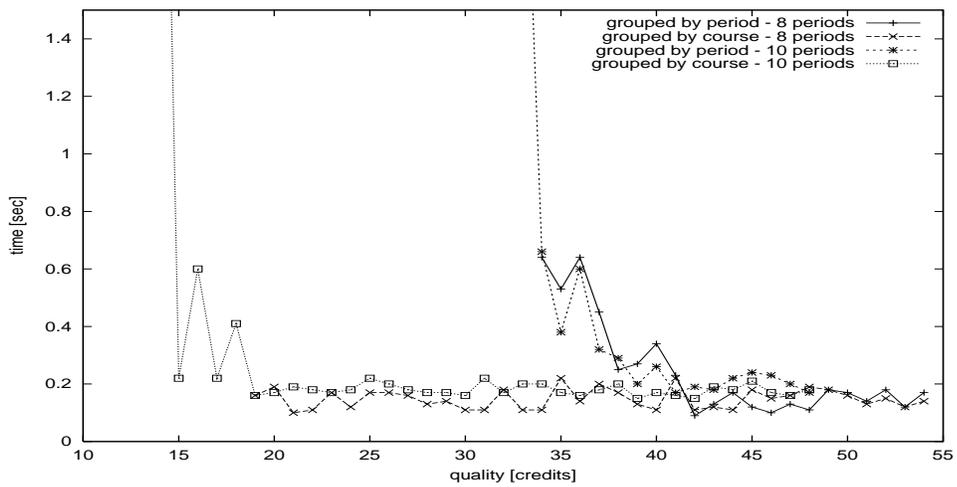,height=6.5cm,width=13cm}
    \caption{Results obtained varying the variable ordering in the
    Oz model}
    \label{fig:oz_8-10_IJ_naive}
  \end{center}
\end{figure}

Inspecting the implementation, we noticed that the way Oz chooses the
next variable to instantiate (left to right) from the binary vector is
relevant to the way we interpretate that vector as a matrix. Varying
then the variable ordering changed the results as shown in figure
\ref{fig:oz_8-10_IJ_naive}.

We got the optimal solution for the 10-period problem, but still have
problems reaching the $8$-period problem optimum\footnote{The optimum is
reached at \texttt{C = 17} in the $8$-period problem, we did not have
any response after $5$ hours for \texttt{C <= 18} with the assignment
vector grouped by course.}. The $12$-period problem reports partial
solutions with \texttt{C >= 28}, but still far away from the optimum.

\begin{figure}[ht!]
  \begin{center}
    \leavevmode
    \epsfig{file=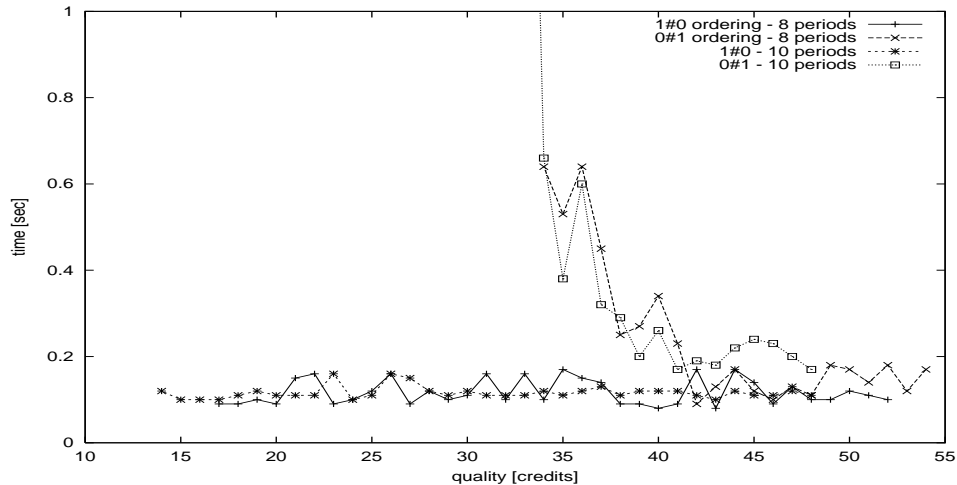,height=6.5cm,width=13cm}
    \caption{Results obtained varying the value ordering in the
    Oz model with the assignment vector grouped by period}
    \label{fig:oz_8-10_JI_reverse}
  \end{center}
\end{figure}

Looking for the reason why Oz could not solve the $8$-period problem, we
tried changing another parameter, the \emph{value ordering}. The way
Oz was instantiating the variables was not natural because it was
first trying the value \texttt{0} which means not to assign the course
to a period. Then it first prohibits any assignment until it gets a
\emph{failed space}, and backtracks searching for solutions. Given
that for instantiating the courses we were using the variable ordering
specified by the University's original academic curriculum, a good
approach seemed to be to assign the courses in that order. Varying
this value ordering we were able to get solutions for any quality
requirement in a constant time\footnote{The fluctuations are due to
machine noise and low precision in the Oz measurement tool used.} and
get the optimum for every problem. The results are presented in figure
\ref{fig:oz_8-10_JI_reverse}.

\begin{figure}[ht!]
  \begin{center}
    \leavevmode
    \epsfig{file=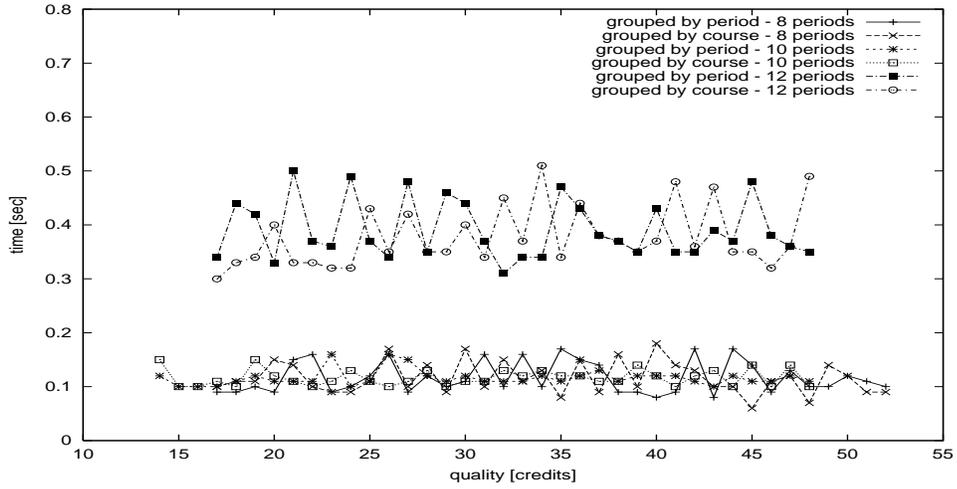,height=6.5cm,width=13cm}
    \caption{Results obtained varying the variable ordering in the
    Oz model with the value order reversed}
    \label{fig:oz_8-12_XX_reverse}
  \end{center}
\end{figure}

Finally, we compared the variable ordering with the new value ordering
and got no significant differences as shown in figure
\ref{fig:oz_8-12_XX_reverse}.

\newpage

\vspace{-14mm}

A comparative summary of the experimental results is given in table
\ref{tab:lp_solve_oz_summary}.

\begin{table*}[ht!]
\center
\begin{tabular}{|l|c|c|c|c|c|}\hline\hline
 \multicolumn{6}{|c|}{8-period problem} \\\hline\hline
 & \texttt{lp\_solve} & \multicolumn{4}{|c|}{Oz} \\ \cline{3-6}
 &                    & \multicolumn{2}{|c|}{grouped by course} &
 \multicolumn{2}{|c|}{grouped by period} \\\cline{3-6}
 &                    & naive & reverse & naive & reverse \\\hline\hline
Time in seconds for first & & & & & \\
solution (worst quality) & 1.7 & 0.1 & 0.1 & 0.2 & 0.1 \\\hline
Time in seconds for optimum & & & & & \\
solution (best quality) & 1459.7 & $\infty$ & 0.1 & $\infty$ & 0.1 \\\hline\hline
 \multicolumn{6}{|c|}{10-period problem} \\\hline\hline
Time in seconds for first & & & & & \\
solution (worst quality) & 5.9 & 0.2 & 0.1 & 0.2 & 0.1 \\\hline
Time in seconds for optimum & & & & & \\
solution (best quality) & $\infty$ & 3.6 & 0.1 & $\infty$ & 0.1 \\\hline\hline
 \multicolumn{6}{|c|}{12-period problem} \\\hline\hline
Time in seconds for first & & & & & \\
solution (worst quality) & $\infty$ & 0.4 & 0.5 & $\infty$ & 0.3 \\\hline
Time in seconds for optimum & & & & & \\
solution (best quality) & $\infty$ & $\infty$ & 0.3 & $\infty$ & 0.3 \\\hline\hline
\end{tabular}
\caption{Comparative summary of performance between \texttt{lp\_solve}
  and Oz}
\label{tab:lp_solve_oz_summary}
\end{table*}

\vspace{-10mm}

All details about the Integer Programming model and the implementation
in Oz of the constraint-based model, as well as the results obtained
using both approaches, can be obtained at {\tt
http://www.labsc.inf.utfsm.cl/\~{}smanzano}.

\section{Related Work}

\vspace{-2mm}

The works by Henz~\cite{Henz:1999} and by Curtis, Smith, and
Wren~\cite{CurtisSmithWren:1999} are the most recent references on the
use of Constraint Programming and Integer Programming for solving
combinatorial optimization problems. Dincbas, Simonis and Van
Hentenryck discuss a case in which the expressive power of Constraint
Programming allows to reformulate an Integer Programming model giving
a much smaller constraint-based model and reducing in this way the
size of the search
space~\cite{DincbasSimonisVanHentenryck:LP:1988}. Van Hentenryck and
Carillon~\cite{VanHentenryckCarillon:AAAI:1988} deal with a warehouse
location problem and Smith, Brailsford, Hubbard and
Williams~\cite{SmithBrailsfordHubbardWilliams:TR-8:1995} apply both
techniques for solving the progressive party problem. In both cases,
Constraint Programming does better than Integer Programming mainly due
to the use of appropriated variables. In our case, we do use the same
variables and the same set of constraints. The very high efficiency of
Constraint Programming was obtained thanks to the suitable heuristics
used for enumerating the variables.

\vspace{-2mm}

\section{Conclusions}
\label{conclusions}

\vspace{-2mm}

We have presented the use of constraint programming techniques for
solving a combinatorial optimization problem. Using constraint
programming we have been able to solve simple problems that cannot be
solved by integer programming techniques. Moreover, we can solve
medium size problems very efficiently. An important variablity can be
obtained when using constraint programming: on one hand, using naive
heuristics we can not solve even simple problems, on the other hand,
using clever heuristics we can solve very efficiently some problems
that are too hard for integer programming techniques. Of course, more
work is needed to understand when to apply each of the heuristics used
in this paper.

\vspace{-3mm}

{\footnotesize

\bibliographystyle{abbrv}


\end{document}